# Dynamics of single human embryonic stem cells and their pairs: a quantitative analysis

L. E. Wadkin[1], L. F. Elliot[1], I. Neganova[2], N. G. Parker[1], V. Chichagova[2], G. Swan[1], A. Laude[3], M. Lako[2], and A. Shukurov[1*]

[1] *School of Mathematics and Statistics, Newcastle University, Newcastle upon Tyne, UK*
[2] *Institute of Genetic Medicine, Newcastle University, Newcastle upon Tyne, UK*
[3] *Bio-Imaging Unit, Medical School, Newcastle University, Newcastle upon Tyne, UK*

[*]*Corresponding author*
Prof. Anvar Shukurov
School of Mathematics and Statistics,
Newcastle University,
Newcastle upon Tyne, UK
Phone: +44 (0) 191 208 5398
Fax: +44 (0) 191 208 8020
e-mail: anvar.shukurov@ncl.ac.uk

**Author contributions**

I.N., N.G.P., M.L. and A.S. designed and supervised the study. L.E.W. and G.S. extracted the data. L.E.W. performed the statistical analysis and prepared the figures. I.N. conducted the experiments. A.L. performed the time-lapse microscopy. L.E.W., I.N., N.G.P., M.L. and A.S. wrote the manuscript. V.C. and L.F.E. contributed to discussions. N.G.P., M.L. and A.S. performed fund raising.

**Competing financial interests**
The authors declare no competing financial interests.

**Abstract**
Numerous biological approaches are available to characterise the mechanisms which govern the formation of human embryonic stem cell (hESC) colonies. To understand how the kinematics of single and pairs of hESCs impact colony formation, we study their mobility characteristics using time-lapse imaging. We perform a detailed statistical analysis of their speed, survival, directionality, distance travelled and diffusivity. We confirm that single and pairs of cells migrate as a diffusive random walk for at least 7 hours of evolution. We show that the presence of Cell Tracer significantly reduces hESC mobility. Our results open the path to employ the theoretical framework of the diffusive random walk for the prognostic modelling and optimisation of the growth of hESC colonies. Indeed, we employ this random walk model to estimate the seeding density required to minimise the occurrence of hESC colonies arising from more than one founder cell and the minimal cell number needed for successful colony formation. Our prognostic model can be extended to investigate the kinematic behaviour of somatic cells emerging from hESC differentiation and to enable its wide application in phenotyping of pluripotent stem cells for large scale stem cell culture expansion and differentiation platforms.

# INTRODUCTION

Human pluripotent stem cells (encompassing both hESCs and the human induced pluripotent stem cells (hiPSCs)) hold great potential for advancement of cellular therapies, disease modelling and drug discovery. Under standard culture conditions hESCs and hiPSCs grow as colonies, and due to the protocols used for their propagation the arising colonies are often characterised by mixed clonal origin. Also, the extensive cell death after enzymatic treatment upon cell passaging results in a very low single-cell cloning efficiency, typically less than 1%[1], even in the presence of the inhibitor of Rho-associated kinase (ROCK)[2]. Moreover, the presence of ROCK was shown to increase cell motility, thus contributing to the development of clones originating from more than one founder cell[3]. Individual cell movement and asymmetric colony expansion negatively impact the accuracy of the hESC clonogenic assays when using a low-density seeding approach with ROCK[5]. This matter highlights the need for a deeper understanding of the processes by which individual hESCs generate pluripotent stem cell colonies. It has been suggested that the local microenvironment modulates the endogenous parameters that can be used to influence hESCs differentiation trajectories[4]. To bring hESCs/hiPSCs differentiation protocols to large-scale assays and into clinical trials, there is a great need for controlled and reproducible cell production strategies. This is a point where understanding of the rules and regulation of pluripotent hESC colonies and their formation from individual cells would benefit.

Single hESCs are reported to undergo an apparently random walk pattern of movement when the cells are more than about 150 $\mu$m apart[3]. The cells that are closer to each other move in a more systematic, directed manner, and display a higher ability to form colonies arising from more than one founder cell, suggesting that the separation distance of hESCs at the start of the colony formation from a few cells and their migration parameters are important for clonal expansion and thus have an impact on pluripotent phenotype and status of a colony as a whole[5]. Equally, the cellular cross-talk is also important when colony growth ensues, as it has been shown that a band of differentiated cells forms at the outer edge of colonies producing an annulus that remains constant in width as the colony grows[5]. This observation suggests important constraints on the proliferation of hESCs and dependence on the cell position within a colony.

In a last few years, several groups have employed time-lapse analysis to study the behaviour of hESCs during the colony formation[6,7]. However, their attention was focused on multicellular colonies rather than early stages of their formation. To date, mechanisms that regulate the heterogeneous phenotype of a hESC colony and its relation to the functional properties of a single cell are poorly studied. The link between the properties and behaviour of individual cells and their fate with respect to the formation of a hESC colony is not well understood and this hampers our ability to predict hESC differentiation capacity under different cell seeding densities.

Thus, our aim is to develop a quantitative model and understanding which would facilitate prediction of the behaviour of individual cells during the colony formation as well as within the colony. This would allow a non-invasive characterization of hESC colonies, identification of cell fate history and their interaction with neighbours. Such a model would provide a useful platform for testing impacts of small molecules, Cell Tracers, differentiation agents and culture media, and enable molecular studies of mutual cell interactions. To approach this major task systematically, we should first understand the behaviour of single hESCs, cell pairs, small clumps and then, growing colonies. At each stage, the aspects of single cell behaviour associated with its neighbouring cells have to be clearly identified and quantified.

Here we focus on the behaviour of single hESCs and their pairs as a first step towards understanding how a hESC colony is formed, by providing a detailed statistical analysis of their kinematic behaviour, including their speed, survival, directionality, distance travelled and division time. In particular, we extend on the previous work[3] by rigorously demonstrating that single cells migrate in accord with a random walk consistent with diffusive motion, and determine their diffusivity, the single parameter which completely specifies the walk (see below for a detailed explanation). It is important to note that the cell motion can only be approximated as a diffusive random walk over a limited time interval since cells inevitably divide, die and interact with other cells, and our results help to identify the extent to which this model behavior holds. We also extend the migration analysis to cell pairs. Mutual interactions of closely positioned cells strongly affect the migration, and we identify two distinct behavioural regimes. Also, the cell pair as a whole is shown to undergo a random walk with characteristic diffusivity. In establishing the diffusive migration of cells and cell pairs, our work opens the possibility to use the well-developed, powerful mathematical theory of random walks for non-invasive prognostic modelling of the behaviour of single hESCs, with obvious implications for large scale expansion and differentiation assays.

We also demonstrate that key characteristics are compromised when the hESCs are stained with Cell Tracer, resulting in lower survival rate, longer time to the first cell division and reduced migration velocity; nevertheless, the diffusive random walk remains an accurate description of the cell migration, thus confirming the validity of our mathematical model and its usefulness for easy and non-invasive assessment of changes in hESC culture environment.

## MATERIALS AND METHODS

### Human embryonic stem cell culture

Human embryonic stem cell line H9 at passages 40-42 (WiCell, Madison WI) was maintained on Matrigel® Basement Membrane Matrix (Corning Inc.) in mTeSR™1 medium (STEMCELL Technologies). Cells were dissociated into single-cell suspension with StemPro® Accutase® (Thermo Fisher) which was diluted by 50% with PBS and plated on 6-well plates pre-coated with Matrigel at a density of 1500 cells/cm$^2$ (as described by Li *et al.*[3]) in mTeSR™1 media supplemented with 10 $\mu$M of Y-27632 (ROCK inhibitor, Chemdea) for the first five hours. Afterwards, cells were fed with TeSR™1 media and divided in two groups: in the absence and presence of Cell Tracer. The former group was incubated with 3 $\mu$M of DMSO for 20 minutes at 37°C. The latter group was stained with 3 $\mu$M CellTrace Violet Dye (Thermo Fisher) according to the manufacturer's protocol. Then, cells were washed twice with culture media before fresh mTeSR™1 was applied. After 1 hour cells were observed under time-lapse imaging. Apart from the opportunity to assess cell division and generate cell lineage trees, this methodology allows us to explore the response of the cell dynamics to the Cell Tracer treatment, and to identify systematic features of the cell behaviour.

**Time-lapse cells video imaging and tracking**

Cells were imaged using a Nikon Eclipse Ti-E microscope, with images recorded every 15 minutes over a total duration of 66 hours for the experiment in the absence of Cell Tracer and 92.25 hours for the experiment performed with Cell Tracer. Each image, analysed within the Nikon NIS-Elements software, had a resolution of 0.62 and 0.96 μm/pixel, respectively. The sampling time of 15 min was chosen to ensure that the typical cell displacement between the images (4–6 μm) is several times the pixel size (such that the pixel discretisation has no significant effect on the trajectories and ensuring that the positions are not over-sampled) and several times smaller than the typical cell diameter (allowing each cell to be individually tracked and for the overall motion on scales larger than the cell size to be adequately sampled). Only cells which had no neighbours within the 150 μm radius were followed in view of the observations by Li *et al.*[3] that the mutual interaction of the cells is negligible at this and larger separations, thus preventing development of colonies arising from more than a single cell. The single cells were tracked by manually recording their centroid coordinates in every frame. This approach is used extensively for cell tracking in general[8], as well as for hESCs, see e.g. Li *et al.*[3]. While a typical single-cell displacement over the course of our experiments is not much larger than a cell diameter (typically a few cell diameters), we expect that our main results are robust to different definitions of the cell position (e.g. nucleus position) since the statistical scatter is itself of a similar size to the cell diameter. Upon cell division, the daughter cells were tracked as a cell pair. Tracking of a single cell ceased when the cell underwent apoptosis; cell pairs were tracked until one of the cells underwent apoptosis or division, or one or both cells could no longer be clearly identified. The number of cells that died was also recorded, along with their time of death.

**Quantification of the cell motion and random walks**

A feature of this work is to establish rigorously the applicability of the isotropic random walk model to the *in vitro* cell migration observed in the experiments. A tortuous, apparently random trajectory of a cell movement does not necessarily imply that it can be described as a random walk (beyond the casual meaning of the phrase), and this should be proved through careful quantitative analysis of the cell movements. Here we introduce the properties of the isotropic random walk, derive the quantitative parameters of the cell migration, and deduce the defining descriptive parameters that can be used for predictive modelling. Several distinctive features characterise the simplest random walk. The migration is isotropic, i.e. there is no preferred direction in the cell movement. It is natural to expect that the migration is isotropic in the absence of large-scale gradients in the environment, and far away from any boundaries. It is then important to establish a quantitative measure of the isotropy in order to detect any deviations from it that may arise from, e.g. inter-cell interactions.

An idealisation involved in the isotropic random walk description is the assumption that a cell moves along a straight line for a short period of time $\tau$, covering a distance $l$, and then changes its direction of migration at random, with any direction having equal probability. This idealisation is adequate with time-lapse imaging where frames are taken at discrete times. We note, however, that the *correlation time* of the random walk, denoted $\tau$, is an intrinsic property of the migrating hESCs, unrelated to the frequency of the image recording. A single summary parameter, that characterises an isotropic random walk completely at long time or space intervals, is known as the *diffusivity D,* such that the root-mean-square displacement traversed in time $t$ is given by

$$\left(\overline{L^2}\right)^{1/2}(t) = \sqrt{2Dt}. \qquad (1)$$

The larger is the diffusivity, the larger is the mean displacement traversed in a given time; we also note that, unlike a directed motion, the displacement traversed in a random walk grows as the square root of time rather than linearly. The diffusivity associated with a two-dimensional isotropic random walk is related to the correlation time and the speed of the cell movement $v$ as $D = \frac{1}{2}\tau v^2$. Also, $l = v\tau$ is the length of a straight leg of the random walk.

Consider a cell, initially positioned at a point with coordinates $(x_0, y_0)$, moving along a certain path in the $(xy)$-plane. After some time $t$, the cell is at a different position $(x(t), y(t))$. The displacement of the cell at a time $t$, measured along the straight line from the starting point is $L_i(t) = \sqrt{[x_i(t) - x_{i,0}]^2 + [y_i(t) - y_{i,0}]^2}$, with $i$ being the cell identifier. We calculate the mean-square displacement from the experimental observations as $\overline{L^2}$, where overbar denotes the average over all cells. As a further confirmation of the diffusive character of the walk, and to draw a link to the work of Li et al.[3], we also consider the directionality of the walk, which, at a given time, is defined as the displacement of the walker divided by its integrated path length up to that time. The directionality is further described in Section 1 of the Supplementary Information.

The movement of a pair of cells can be characterised by two distinct velocities. The first of these is the velocity of the pair as a whole, i.e. its pair centroid velocity $\mathbf{v}_{pc}$. Denoting the individual velocities of the cells in the pair as $\mathbf{v}_1$ and $\mathbf{v}_2$, then the pair centroid velocity is the vectorial average of these, i.e. $\mathbf{v}_{pc} = (\mathbf{v}_1 + \mathbf{v}_2)/2$. The second of these quantities is the relative velocity of the cells within the pair $\mathbf{v}_r$, defined as the difference of the individual velocities, $\mathbf{v}_r = \mathbf{v}_1 - \mathbf{v}_2$. This velocity characterises the approach of the cells or their motion away from each other. From these two quantities we define the corresponding speeds as the vector magnitudes, $v_{pc} = |\mathbf{v}_{pc}|$ and $v_r = |\mathbf{v}_r|$. It can be expected that the relative velocity of the cells in a pair can no longer be approximated by an isotropic random walk because of their mutual interaction. However, the pair as a whole still might perform a random walk, but its diffusivity can be different from that of a single cell.

We measured positions of individual cells in the time sequences of microscope images, with special attention to the cross-identification of each cell tracked in subsequent images and to cell division; the cells produced in each division were labelled as being related. The resulting array of cell positions was imported into Matlab for further analysis, which was performed using matrix operations.

Lineage trees were constructed using Matlab by extracting, for each cell, the time to first division or death, and marked appropriately. Time to second division for pairs was also extracted. Tracking of cells trajectories is straightforward with Matlab as we have a matrix of each cell's positions at each timeframe. In general we use the median to characterise representative values of variables (since the sample sets are not always normally distributed), with the errors representing the lower and upper quartiles. The exception is the root-mean-square displacement $\left(\overline{L^2}\right)^{1/2}$, which is quoted in terms of means (according to its definition).

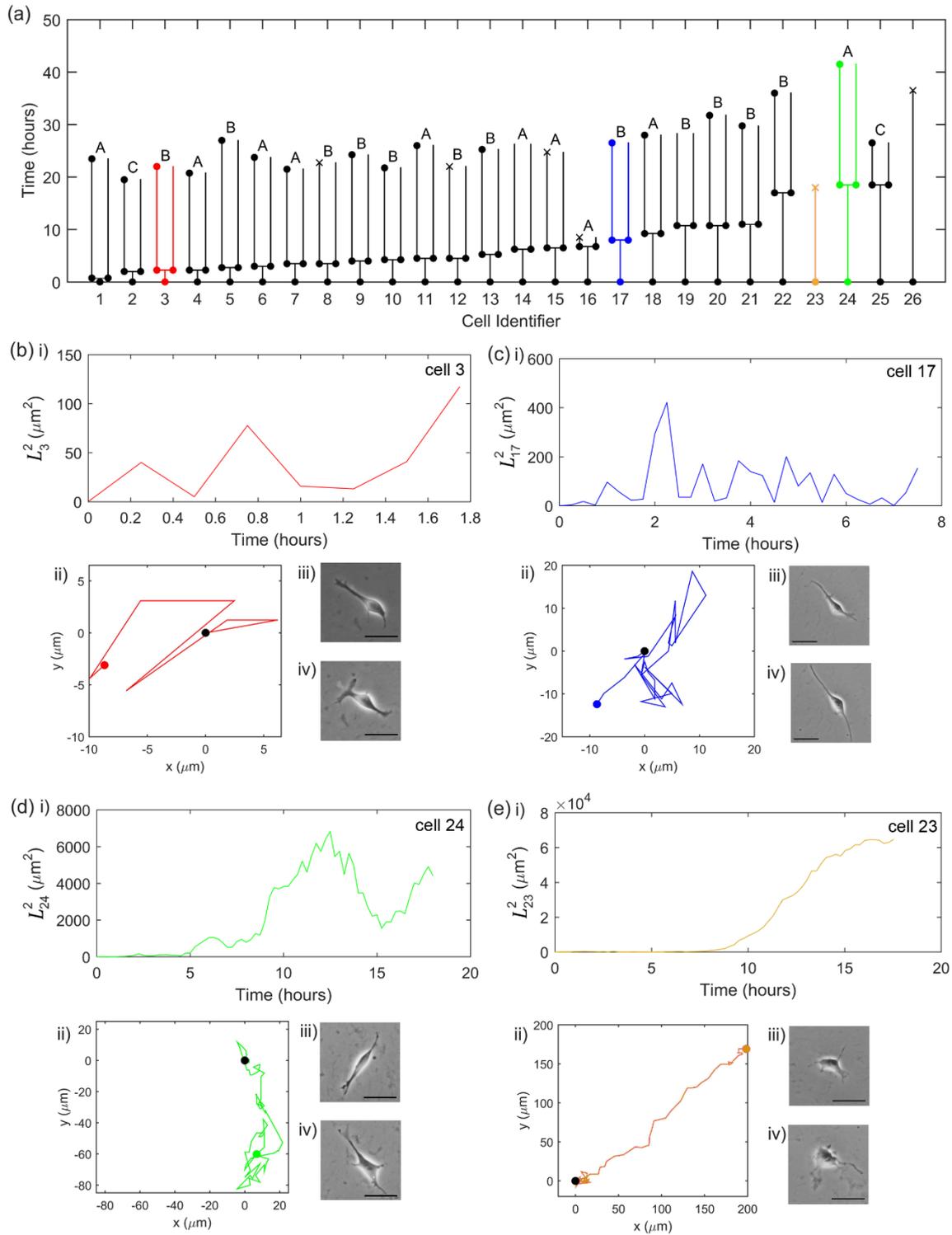

Figure 1: Single-cell behaviour in the absence of Cell Tracer. (a) The timelines of the 26 single cells. A filled circle indicates that the cell has divided, a cross indicates the cell has died, and no marker indicates that the image was not clear enough to identify the cell confidently; in both cases, we did not track cells beyond this point. The range of behaviours is illustrated with Cell 3 (red), 17 (blue), 23 (orange) and 24 (green), as detailed in Panels (b)–(e). The predominant behaviour is that similar to Cell 17. In each case, we show (i) $L_i^2$ as a function of time, (ii) the cell trajectory (with the black and coloured circles indicating the start and end of the trajectory, respectively), and its microscopy images at (iii) the start of the recording and (iv) close to the end of its walk. The length of the scale bar shown within the microscopic images is 50 $\mu$m. The sampling interval is 15 minutes and the microscope resolution is 0.62 $\mu$m/pixel.

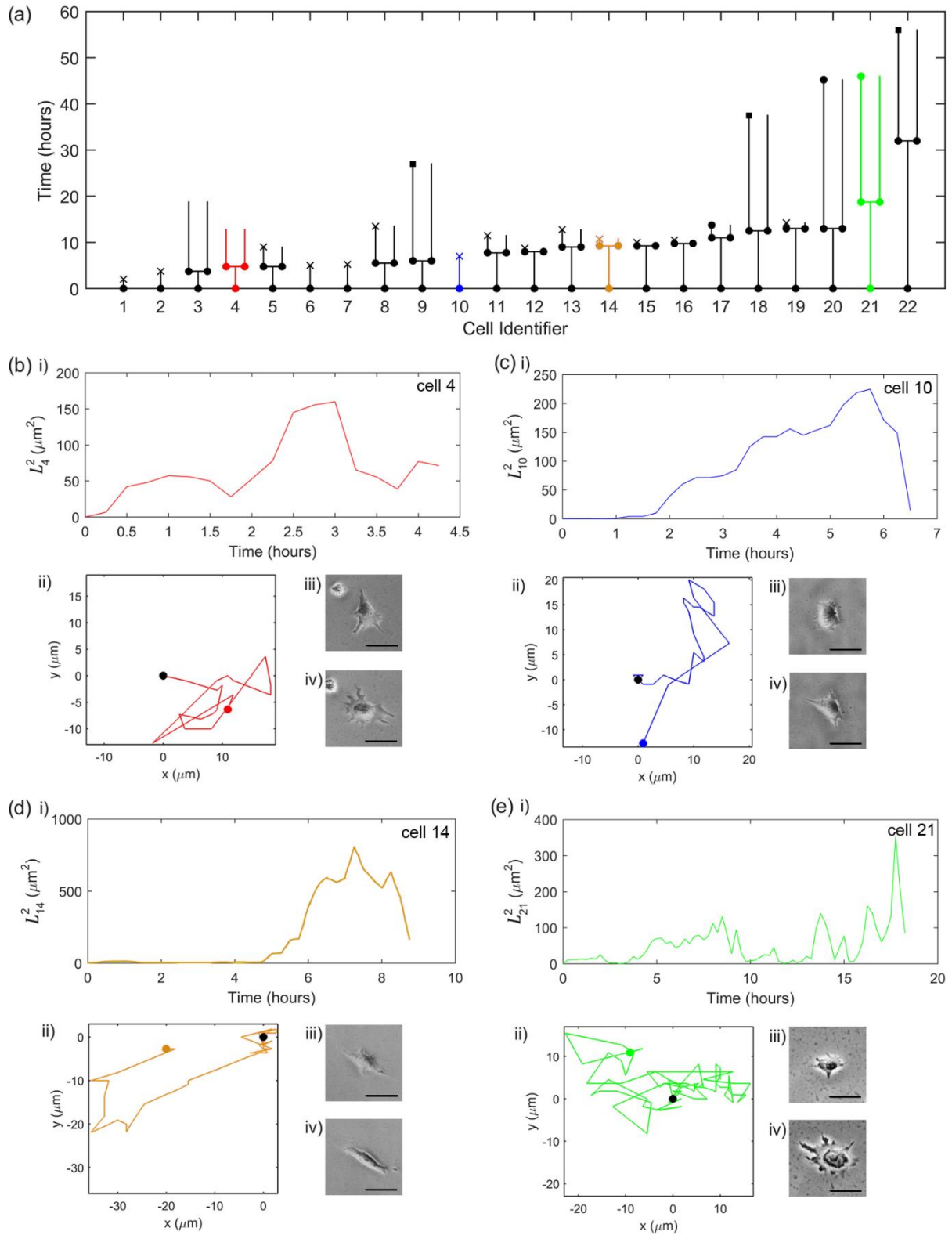

Figure 2: Single-cell behaviour after staining with the Cell Tracer. The panels are as in Figure 1, but Cells 9, 18 and 22 are marked with a square in Panel (a) to indicate that they eventually join a larger colony and are not traced after that. The predominant behaviour is that similar to that of Cell 4. The microscopy image resolution is 0.96 $\mu$m/pixel.

# RESULTS

## Kinematics of single hESCs

To analyse the behaviour of single hESCs, both in the absence and in the presence of Cell Tracer dye, we identified and tracked the initially isolated cells in each experiment. To enable single cell identification, we built lineage trees, shown in Figure 1a in the absence of Cell Tracer and Figure 2a with Cell Tracer, in which cells were ordered according to their time to the first division (or death) and given an identification number. From the tracked single cells we calculated various quantities that characterise their movement, including their speeds, their mean-square displacements from initial positions, correlation times and diffusivities, as well as other parameters such as death rate and time to first division. The main quantities extracted from the data are listed in Table 1. Corresponding mean values for the parameters are given in Table T1 in Supplementary Information. Below we will discuss these measurements in detail, first for the absence of Cell Tracer, and then in the presence of Cell Tracer.

Table 1. Summary of parameters acquired for single cells cultured in the absence and presence of Cell Tracer. The entries are representative median values, with errors given by the lower and upper quartiles. The exception is the diffusivity, which is presented as a mean and 95% confidence interval. The step lengths and migration speed were calculated by averaging over all cells at all times (15 min intervals). The diffusivity was obtained using the fits to $\bar{L^2}$ shown in Figure 3 and the correlation time from $\tau = 2D/v^2$ for instantaneous speeds.

| Parameter and notation | | No Cell Tracer | With Cell Tracer |
|---|---|---|---|
| Number of cells | $N$ | 26 | 22 |
| Migration speed ($\mu$m/hr) | $v$ | $16.25^{+8.4}_{-3.9}$ | $11.51^{+3.3}_{-3.8}$ |
| Step length in $x$ ($\mu$m) | $l_x$ | $2.5^{+2.5}_{-1.3}$ | $1.9^{+1.0}_{-0.9}$ |
| Step length in $y$ ($\mu$m) | $l_y$ | $2.5^{+2.5}_{-1.3}$ | $1.9^{+1.0}_{-0.9}$ |
| Correlation time (hr) | $\tau$ | $0.6^{+0.9}_{-0.4}$ | $0.7^{+1.1}_{-0.5}$ |
| Diffusivity ($\mu m^2$/hr) | $D$ | $79.8 \pm 5.2$ | $49.1 \pm 3.5$ |
| Time to first division (hr) | $t_{\text{div}}$ | $4.9^{+5.1}_{-1.6}$ | $9.3^{+3.3}_{-3.4}$ |
| Death fraction at 10 hours, % | – | 0 | 23 |
| Death fraction at 20 hours, % | – | 4 | 23 |

*Absence of Cell Tracer*

In the absence of Cell Tracer, we followed 26 isolated cells, with their summary behaviour presented in Figure 1. The lineage tree, shown in Figure 1a, enables cell identification with the cells ordered according to their time to the first division (or death) and given an identification number from 1 to 26. Of these, 24 divided, with this first division, on average, occurring at $4.9^{+5.1}_{-1.6}$ hours (see Supplementary Figure S3 for a histogram of the time for first division). The remaining two cells died, Cells 23 and 26. Of the 24 surviving pairs, 20 went on to a further division, and in the remaining four pairs, one of the cells died.

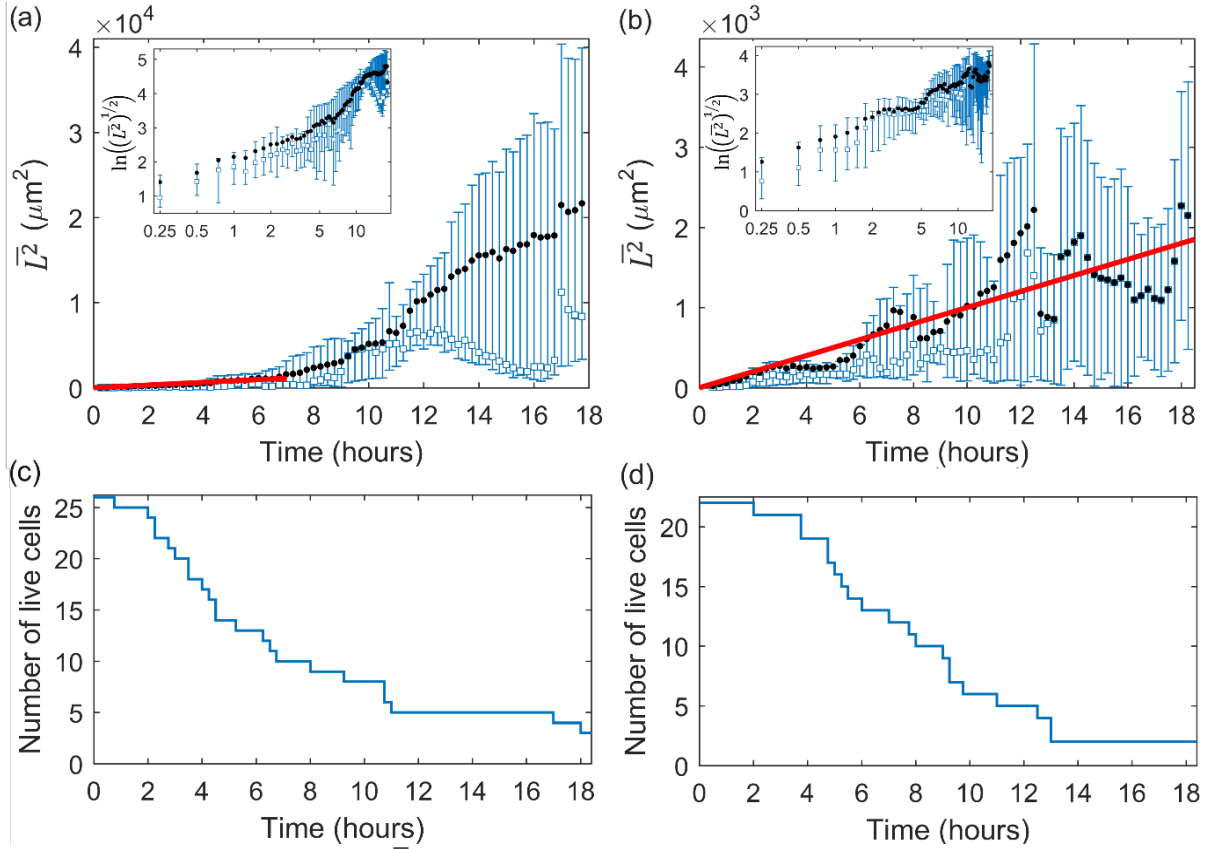

Figure 3: The evolution of the mean $\bar{L}^2$ (black circles) and median (blue squares) displacements of single unstained cells (a) and stained cells (b) with red lines showing straight-line least-squares fits (constrained to pass through the origin) of $\bar{L}^2 = 2Dt$ with $D = 79.8 \pm 5.2$ $\mu$m$^2$/hr for unstained cells and $D = 49.1 \pm 3.5$ $\mu$m$^2$/hr for stained cells. Error bars show the upper and lower quartiles. Insets show the root-mean-square displacement $\left(\bar{L}^2\right)^{1/2}$ vs time on natural logarithmic axes. The number of live cells over time for unstained cells (c) and stained cells (d) indicate the changing sample size. The sampling interval is 15 minutes.

To illustrate the range of cell behaviours, we focus on four cells: three that divided at different stages during their migration (Cells 3, 17 and 24, shown in Figures 1b, c and d, respectively) and a cell that died before it could divide (Cell 23, shown in Figure 1e). For each case we show the cell's mean-square displacement from its original position over time, its trajectory, and its appearance at the start and end of its walk. Careful examination of morphological images shows an increase of single cell size (30.1 ± 1.5 $\mu$m), also often with multiple filopodia, when compared to cells cultured at high density or as colonies (9.5 ± 3.5 $\mu$m), in agreement with previously published results[3,7]. Note that these four cases of Figure 1 (b-d)ii do not occur at the same frequency; the predominant cell behavior is akin to that of Cell 17.

For Cell 23 (Figure 1e), the migration behaviour is distinct. For the first approximately 7 hours of its migration, this cell moves akin to an isotropic random walk, as seen in Figure 1e(ii). However, after this time the cell moves in an almost straight path. This directed movement persists until the cell dies at around 18 hours. This peculiar behaviour could be due to reasons specific to this particular cell, and we note that the other cell that died after a sufficiently long time interval, Cell 26, behaved in a more typical random walk manner.

To verify if the cell migration is consistent, on average, with the theory of isotropic random walk we consider $\bar{L}^2$ (averaged over the 26 single hESCs) versus time, shown in Figure 3a. From this it is evident that the behavior of $\bar{L}^2$ is approximately linear for around the first 7 hours, before the character changes. The least-squares straight-line fit over this seven hour period (constrained to pass through the origin) is $\bar{L}^2 = (159.6 \pm 10.4)t$, with $\bar{L}^2$ in $\mu m^2$ and $t$ in hours, giving the estimate of diffusivity $D = 79.8 \pm 5.2\ \mu m^2/hr$ from Equation (1). Here and elsewhere, the parameter ranges obtained from the least-squares fits represent 95% confidence intervals. Note that the number of samples (single cells) decreases over time due to cell death and division, and is plotted in Figure 3c. The diffusive behavior over this early period is also confirmed by plotting the average directionality versus time, presented in Supplementary Section S1.

We conjecture that this change in mobility is related to the cell division. The typical time to the first division is about 7 hours (Table 1); thus, by around 6 hours, a significant number of cells have undergone division. This process is known to trigger different cell secretions, which will modify the chemical environment of the cell and so might also affect their kinematic behaviour.

To confirm that the cell migration is isotropic, we consider the step lengths in the two orthogonal directions, $x$ and $y$, per 15 minute frame in the microscopy imaging. There is strong scatter in the individual step lengths, but the average values, $l_x$ and $l_y$, are well defined, and, importantly, are the same in each direction (within statistical error). The values are $l_x = l_y = 2.5^{+2.5}_{-1.3}\ \mu m$. Both the Kolmogorov-Smirnov and Mann-Whitney U test indicate there is no evidence to reject the null hypothesis that the distributions of $l_x$ and $l_y$ are the same. The Pearson product-moment correlation coefficient of $l_x$ and $l_y$ is as small as 0.22, confirming the steps in the $x$ and $y$ directions are uncorrelated; this is further illustrated in Supplementary Figure S4. This confirms the isotropic nature of the cells' random walk.

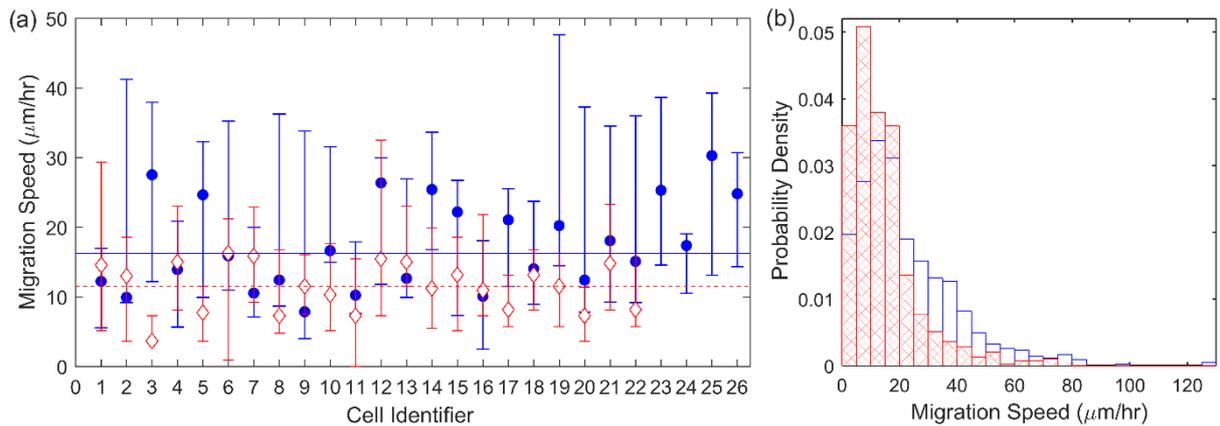

Figure 4: (a) The median migration speed of the individual hESCs in the absence (filled blue circles) and presence of Cell Tracer (open red diamonds). The minimum number of measurements for each cell is 8. Error bars correspond to the upper and lower quartiles. The median migration speeds across all cells, are shown by the blue solid line and red dashed, respectively. Cells are ordered by time to division/death, as per Figure 1. (b) Normalised histogram (probability density) of instantaneous migration speeds for cells in absence (blue) and presence of Cell Tracer (red, crosshatched).

Next we consider the migration speed of the single cells. Figure 4a (blue data) shows the average speed for each cell. It is remarkable that the speeds of all cells are the same within their errors. The average speed across all the cells (without Cell Tracer) is $16.25^{+8.4}_{-3.9}$ $\mu$m/hr. From the estimated diffusivity and the instantaneous speeds of the walks, the average correlation time was determined as $\tau = 0.6^{+0.9}_{-0.4}$ hours. We also consider the spread of migration speeds across the cells, shown in Figure 4b. This also shows a systematic and well-defined trend on a skewed, Maxwellian-like distribution, with low occurrence of very low speeds, a high occurrence of intermediate speeds, and a progressively lower occurrence of increasingly high speeds. Note that the cells in Figure 4a are ordered according to the time to division/death (see Figures 1 and 2). The lack of any noticeable trend in the migration speed across the cells thus ordered indicates that there is no noticeable correlation between the cell migration speed and the time to division or death.

*Presence of Cell Tracer*

In the presence of Cell Tracer, 17 out of 22 single cells survived. Of these 17 pairs, half underwent a further division. The average time to first division is $9.3^{+3.3}_{-3.4}$ hours (see Supplementary Figure S3 for a histogram of the time for first division), which is considerably longer than the case with no Cell Tracer. The higher number of deaths and reduced number of divisions compared to the case of no Cell Tracer is highly indicative of higher apoptosis and lower cell division potential, which was corroborated by flow cytometric analysis indicating a two-fold increase in the percentage of apoptotic cells in presence of Cell Tracer (data not shown).

The four illustrative cell migrations shown in Figure 2b-e show generally similar behavioural patterns for dividing (Cells 4, 14 and 21 in Figure 2) and the cell that dies (Cell 10 in Figure 2) as for the case of no Cell Tracer.

The mean-square displacement over time in the presence of Cell Tracer, shown in Figure 3b, also indicates systematic nearly linear increase over time, characteristic of an isotropic random walk. The least-squares straight-line fit is $\overline{L^2} = (98.2 \pm 7.0)t$, with $\overline{L^2}$ in $\mu$m$^2$ and $t$ in hours, giving the estimate of diffusivity $D = 49.1 \pm 3.5$ $\mu$m$^2$/hr. The fit agrees well with the measurements across the whole observation time (see line in Figure 3b). The diffusive-like behavior is also confirmed by considering directionality over time, as shown in Figure S2.

The step lengths, $l_x = l_y = 1.9^{+1.0}_{-0.9}$ $\mu$m are again identical, with a correlation coefficient of 0.28, further confirming the isotropy of the migration (Figure S4). Both the Kolmogorov-Smirnov and Mann-Whitney U test indicate there is no evidence to reject the null hypothesis that the distributions of $l_x$ and $l_y$ are the same. Notably, however, these lengths are reduced by around 25% in comparison with the cells untreated with Cell Tracer. Since the treatment does not affect the correlation time of the random walk, it must be the average speed of the cell migration that is affected by the treatment, leading to a reduced diffusivity.

In the presence of Cell Tracer the median migration speed across all cells is $11.51^{+3.3}_{-3.8}$ $\mu$m/hr, around a 30% reduction of the speed in the absence of Cell Tracer. The representative speeds of all cells (red data in Figure 4a) again show a remarkably consistent behaviour, lying within their error bars. Figure 4b clarifies the effect of the Cell Tracer on the cell speeds: this treatment significantly reduces the fraction of cells that move at a speed in the range 20–60 $\mu$m/hr and increases the abundancy of those moving at speeds below 20 $\mu$m/hr. A Mann-Whitney U test rejects the null hypothesis that the speeds from unstained and stained groups are samples from continuous distributions with equal medians ($p < 0.01$), confirming that there is a systematic difference between the speed of unstained and stained cells. From the estimated diffusivity and the instantaneous speeds of the walk, the average correlation time was determined as $\tau = 0.7^{+0.5}_{-1.1}$ hours.

To summarise, the migrations of individual cells can be described, in statistical terms, as an isotropic random walk, independently of the treatment with Cell Tracer. This description is accurate as long as the cells are separated by distances of about 150 $\mu$m, but then their interactions become significant. The migration of the interacting cell pairs is discussed in the next section.

The Cell Tracer reduces the typical cell speed by approximately 30%, from 16 $\mu$m/hr to 11 $\mu$m/hr, but the typical time scale $\tau$ between significant changes in the direction of migration (the correlation time) remains unchanged.

**Kinematics of hESC pairs**

Having characterised and quantified the kinematics of single hESCs, we now extend our analysis to pairs of hESCs. Specifically, we consider pairs that arise from division. The quantitative results are summarised in Table 2. Corresponding mean values for the parameters are given in Table T2 in the Supplementary Information. Below we describe the results in detail, again first for the case in the absence of Cell Tracer and then in the presence of Cell Tracer.

Table 2. Parameters characterising the migration of hESC pairs, both in the absence and in the presence of Cell Tracer. We present the representative median values, and errors as their upper and lower quartiles. The diffusivity was obtained for Type A cells using the fit to $\overline{L^2}$ shown in Figure S7a and the correlation time from $\tau = 2D/v^2$ for instantaneous speeds.

| Parameter and notation | | No Cell Tracer | | With Cell Tracer |
|---|---|---|---|---|
| | | **Type A** | **Type B** | |
| Number of pairs | $N$ | 10 | 12 | 17 |
| Tracking time (hr) | | 23 | 24.25 | 31.5 |
| Initial separation ($\mu$m) | $s_0$ | $16^{+2.5}_{-1.5}$ | $17^{+5.7}_{-2.4}$ | $16^{+6.2}_{-2.4}$ |
| Final separation ($\mu$m) | $s_f$ | $91^{+81.9}_{-32.3}$ | $34^{+12.6}_{-14.2}$ | $19^{+4.3}_{-4.8}$ |
| Maximum separation ($\mu$m) | $s_{max}$ | $95^{+88.9}_{-14.2}$ | $98^{+52.1}_{-19.0}$ | $26^{+7.7}_{-5.0}$ |
| Time to $\bar{s}_{max}$ (hr) | $T_{max}$ | $17^{+2.6}_{-4.4}$ | $8^{+2.8}_{-3.9}$ | $4^{+5.3}_{-2.3}$ |
| Speed of each cell ($\mu$m/hr) | $v$ | $20.0^{+13.37}_{-8.9}$ | $20.5^{+11.9}_{-8.1}$ | $15.4^{+7.2}_{-6.8}$ |
| Pair centroid speed ($\mu$m/hr) | $v_{pc}$ | $15.1^{+9.0}_{-6.1}$ | $16.2^{+8.8}_{-7.1}$ | $11.7^{+6.0}_{-4.8}$ |
| Relative speed ($\mu$m/hr) | $v_r$ | $31.1^{+17.1}_{-12.2}$ | $27.7^{+17.0}_{-11.8}$ | $16.3^{+10.9}_{-7.7}$ |
| Pair centroid correlation time (hr) | $\tau$ | $0.5^{+0.8}_{-0.3}$ | - | - |
| Pair centroid diffusivity ($\mu$m$^2$/hr) | $D$ | $58.5 \pm 1.8$ | - | - |

*Absence of Cell Tracer*

In the absence of Cell Tracer we consider the 24 cell pairs from the lineage tree (Figure 1a).

Of these, 2 cases could not be characterised because of poor microscope images. The migration behaviour of the remaining 22 co-lineage cell pairs can be classified into two distinct categories, as illustrated in Figure 5:

**Type A** (10 pairs): The cells appear to constantly repel each other (Figure 5a, c), and their separation $s$ increases steadily with time. The average separation of the cells in those pairs at the end of tracking was $(123 \pm 27)$ $\mu$m, equal within errors to their average maximum separation, $(141 \pm 30)$ $\mu$m. These distances are several times the typical cell diameter.

**Type B** (12 pairs): The cells initially appear to repel each other as shown in Figure 5b, d, but then their separation reduces. The average maximum separation of the cells within such pairs is $(117 \pm 17)$ $\mu$m, which again is several times the typical cell diameter, whereas they are located just $(42 \pm 9)$ $\mu$m apart at the end of the tracking.

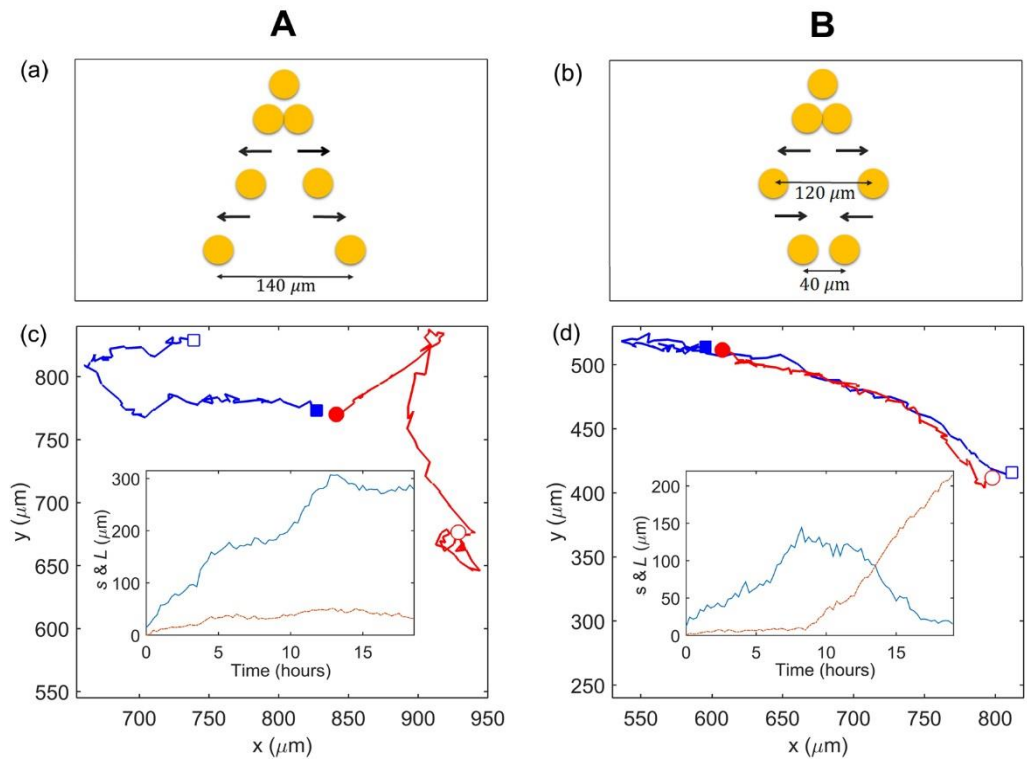

Figure 5: (a) Typical behaviour of a Type A pair, in which the cells appear to repel continually over time, and (b) of a Type B pair, where the cells first appear to repel and then attract each other. (c) The trajectories of the cells in a representative Type A pair (originating from Cell 14 in Figure 1a), with the initial (final) centroid positions marked with filled (open) symbols. Inset: the separation between the cells, $s$ (blue line) and the displacement of the pair's pair centroid position (orange line). (d) As in (c), but for a representative Type B pair (originating from Cell 13 in Figure 1a). The cells are shown as circles or squares only for illustrative purposes.

The median migration speed of individual cells in a pair, $v$, for Type A is $20.0^{+13.37}_{-8.9}$ $\mu$m/hr, while Type B cells move at a speed $20.5^{+11.9}_{-8.1}$ $\mu$m/hr. The speeds are similar for Types A and B, but higher than the speed of a single cell (Table 1).

The difference between the final separations, the defining distinction of Type A and B pairs, is statistically significant ($p < 0.01$) according to the Mann-Whitney U test.

The median pair centroid speeds $v_{\text{pc}}$ for Type A and Type B pairs are $15.1^{+9.0}_{-6.1}$ μm/hr and $16.2^{+8.8}_{-7.1}$ μm/hr, respectively, and the median relative speeds $v_{\text{r}}$ are $31.1^{+17.1}_{-12.2}$ μm/hr and $27.7^{+17.0}_{-11.8}$ μm/hr. These estimates show that the pair as a whole moves at a similar speed to a single cell, but the cells move relatively fast within the pair. The average pair centroid and relative speeds for each cell are shown in Supplementary Figure S5, together with the probability density distributions of the speeds, suggesting that the centroid speeds are similar for Type A and B pair (indeed the Kolmogorov-Smirnov test and Mann-Whitney U test confirm that the speeds are not significantly different). Type B cells eventually move farther away from each other, so it is not surprising that their relative speed is higher than in Type A pairs (Table 2). The Kolmogorov–Smirnov test gives a negative result across the Type A and B relative speeds, confirming that the speeds for these two pair types are from different probability distributions. This is also confirmed by a Mann-Whitney U test for both relative speeds and centroid speeds with $p < 0.01$.

We consider $\overline{L^2}$ over time for pair centroid motion for Type A and B cells, shown in Supplementary Figures S7a and S7b, respectively. For Type A cells this shows that the motion of the pair centroid, i.e. the pair as a whole, can be described as an unbiased random walk with an estimated diffusivity $D = 58.5 \pm 1.8$ μm$^2$/hr (25% less than for single cells) and correlation time $\tau = 0.5^{+0.8}_{-0.3}$ hours. Type B cells exhibit the characteristic random walk behaviour up to around 10 hours (with a similar diffusivity to Type A cells) but not at later times, when their mean square displacement grows faster than diffusive behaviour.

*Presence of Cell Tracer*

In the presence of Cell Tracer we consider the 17 pairs formed from division in the lineage tree of Figure 2a. We are unable to classify the pair dynamics into Type A and Type B; they do not fall conclusively into either category. The median pair centroid and relative speeds for each cell are shown in Supplementary Figure S6. The cell migration speed is $15.4^{+7.2}_{-6.8}$ μm/hr, the pair centroid speed is $11.7^{+6.0}_{-4.8}$ μm/hr, and the relative speed is $16.3^{+10.9}_{-7.7}$ μm/hr. All of these are significantly lower than without Cell Tracer, in qualitative agreement with our observations of the effect of Cell Tracer on single cell migration. The cell speed is slightly higher than that of single cells in the presence of Cell Tracer; this is also consistent with the increased speed of paired cells observed in the absence of Cell Tracer. The evolution of $\overline{L^2}$ for the pair centroids, shown in Supplementary Figure S7(c), shows a linear diffusive behavior for up to around 7 hours, but becomes erratic after that, with no apparent systematic behavior.

To summarize, cell pairs as a whole move akin to an isotropic random walk, but with a lower diffusivity than for single cells. Type A pairs follow this behavior for the whole duration considered, whereas for Type B pairs and Cell Tracer pairs the behavior deviates significantly after around 7 hours. More experimental studies and further modelling are required to clarify the significance and biological cause of this longer-term behavior. The presence of Cell Tracer systematically reduces the speeds of both the pairs and individual cells, consistent with what we observe for single cells. The cells in the pair move slightly faster than in isolation, both with and without Cell Tracer. Perhaps this can be attributed to the more favorable environment in stem cell groups rather than isolation, consistent with the general observation that clustering favors the survival of stem cells[2].

## DISCUSSION

Our analysis of single, isolated hESCs shows that their average migration behaviour is diffusive, akin to an isotropic random walk. Deviations from this behavior are observed after around 7 hours in our experiments; we attribute this to kinematic changes are the cell begin the process of division. Individual trajectories may show significant deviation from the average isotropic random walk behaviour, such as sporadic directed motion, consistent with the probabilistic nature of the individual trajectories. Even with our sample size (of the order of 20, for each experiment, and comparable to that of Li *et al.*[3]), the trend in behaviour is evident. The typical cell displacements we analyse are up to several times the typical cell diameter, which is comparable to that considered in other studies of cell migration[3,9].

In demonstrating the diffusive random-walk-like behaviour of the cells, our work opens the possibility to use the well-established mathematical theory of random walks and diffusion to help plan and optimise experiments with specific aims. For example, agent-based models, which combine the diffusive motion of cells with their biological state and interactions, have strong predictive power for monolayer cultures, as demonstrated for epithelial cells[10]. The quantitative characterisation of single and paired stem cells, as performed here, is a first step to develop these computational technologies towards stem cell cultures.

Moreover, this mathematical framework, coupled with the quantitative data obtained here, allows for the immediate estimation of useful migration-based effects. For example, consider the growth of single-cell clone colony from single cell. It is important to minimise the occurrence of colonies arising from more than one founder cells. Our framework allows us to estimate the typical timescale and probability, from a given seeding density, of this occurrence. For a seeding density $n_0$, the average inter-cell distance is $d_0 = 1/\sqrt{n_0}$. Let us focus on two neighbouring cells separated by this distance, as shown in Figure 6. From their initial positions, each isolated cell performs a random walk with diffusion coefficient $D$. At time $t$, the typical displacement of each cell is $L = \sqrt{2Dt}$, i.e. it lies somewhere on a circle of radius $L$ centered on its initial position. If $L \sim d_0/2$ the circles just touch, indicating that the cells may meet. This occurs at a time $t_0 = d_0^2/(8D) = 1/(8Dn_0)$. This is the typical timescale over which the single cells will meet with a neighbour. Taking, for example, a seeding density of $n_0 = 1500$ cm$^{-2}$ and $D = 80$ $\mu$m$^2$/hr, we obtain $t_0 = 104$ hours.

At this time, if the cells enter within a critical distance $d_c$ of each other (green hatched region) we assume that they will attract and form an aggregate (otherwise they remain unaware of each other). The hatched region subtends an angle $\theta = 2\arccos(d_c/d_0)$ from the initial cell position. Since each cell can be at any point on the circle with equal probability, the probability that the cell falls within the range $\theta$, i.e. the hatched region, is $P = \theta/2\pi$. Since the two cells move independently, the probability that they are both lie in the hatched region is $P^2$. Thus, the probability for the two cells to form a pair at a time $t \sim t_0$ is $\arccos^2(d_c/d_0)/\pi^2$. To keep this probability below a certain value $\alpha$ requires $d_0 > d_c/\cos(\pi\sqrt{\alpha})$ or, equivalently, $n_0 < \cos^2(\pi\sqrt{\alpha})/d_c^2$. This provides an estimate of an optimal seeding density in a cell culture that is grown to produce single-cell clone hESCs colonies, starting at a low density. Based on the above parameters and setting $\alpha = 0.1$ gives $n_0 < 1300$ cm$^{-2}$.

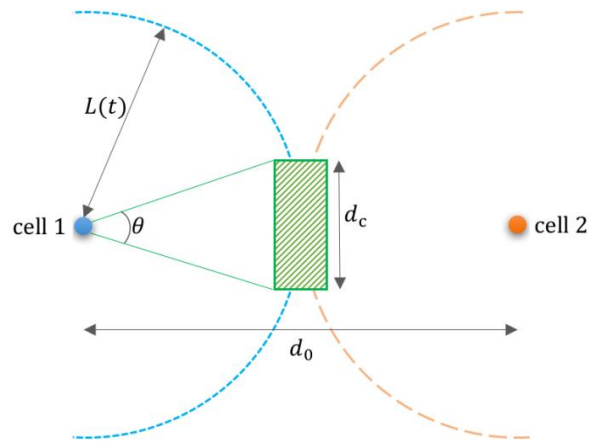

Figure 6: An illustration of the calculation of the probability for two cells to be a distance $d_c$ or less apart at a time $t$ when they perform random walks from initial positions separated by a distance $d_0$. Blue and orange dots show the initial positions of the cells. After a time $t$, each cell is located at a distance $L$ from its original position at some point on the circle shown short-dashed (red) and long-dashed (blue), respectively. To be within a distance less than $d_c$, the cells need to be anywhere in the green rectangular region.

The diffusivity that characterises the random walk behaviour appears sensitive to the factors secreted by the cells. Single and paired hESCs which have been stained with Cell Tracer have significantly lower diffusivities. We attribute this to the negative impact of the Cell Tracer on the cell health and viability, which directly results in less active cell migration. Cell Tracers are widely used in stem cell research and other fields, our research clearly shows that their application should be carefully considered especially when it comes to clonal or low cell density assays in which cell tracers are most likely to have a negative impact.

Remarkably, hESC pairs also move (as a whole) according to an isotropic random walk during about 7 hours of their evolution, albeit with a significantly reduced mobility (the diffusivity of pairs is approximately two-thirds of the single-cell diffusivity) compared to single hESCs. This in turn points to a longer timescale for pairs to encounter other pairs; for example assuming the same plating density parameters as above, the corresponding timescale for an encounter increases to around 140 hours (compared to 104 hours for single cells). This also suggests that the mobility and diffusivity of aggregates of three, four and more cells will be progressively lower, as one would expect intuitively. Indeed, one can extrapolate the relationship between the diffusivity and cell number in the absence of Cell Tracer, and find that this predicts the diffusivity to become zero for an average of around five cells. This roughly agrees with the observation that the smallest group that can develop into a colony contains about five cells[11]. This also implies that the reduction in the diffusivity for the triplets and quadruplets decreases less than between single cells and pairs. This could be tested in future with further experiments. An intriguing observation is the presence of two distinct types of cell pair dynamics (the cells within a pair either continually repel, or at first repel and then attract each other). In future, we plan to study whether these behaviours are related to the presence of different hESC sub-populations, as has been detected elsewhere through the distinct levels of transcription factors across hESC sub-populations[12,13,14].

Our findings suggest that the kinematics of hESC pairs appears to be universal in terms of the cell and pair speeds. Our results agree with the measurements of the speeds of individual hESCs by other researchers that also used a low seeding density of 1500 cells/cm$^2$ [7]. The speed obtained by these authors are about 25, 34 and 24 $\mu$m/hr for their Type 1–3 cells (those that form a colony, join a colony and fail to form a colony, respectively); within the errors, these values are statistically indistinguishable. Moreover, fibroblasts were found to have a significantly higher average speed of about 47 albeit with a high scatter of 60 $\mu$m/hr, so that the differences between all these speeds are perhaps statistically indistinguishable. We confirm that the hESC pairs have kinematic characteristics similar to those of individual cells. Thus, the kinematic diagnostics, if confirmed to be different for pluripotent cells and somatic cells, may lead to rapid, non-invasive diagnostics of the cell pluripotency and differentiation in vitro. This aspect of the hESC and somatic cell kinematics will be explored in a separate publication.

Recent studies using high-throughput single-cell gene expression profiling have uncovered a high degree of cell-to-cell variability within pluripotent stem cell populations. Importantly, regardless of high level of cell-to-cell variability on the level of other parameters (gene and protein expression) we identify parameters of a single cell movement and behaviour, which appears to be true for pairs, and presumably, small groups of hESCs. We used these parameters for prediction of cell seeding density which would allow avoiding appearance of hESCs colonies arising from more than one founder cell due to the aggregation with neighbouring cells and also estimate the optimal timing for achieving clonal expansion at the best confluences. Moreover, our work may be extended to hiPSCs. Given the great similarities in cell cycle progression and pluripotent phenotype between hESCs and hiPSCs, we expect them to exhibit similar kinematic behavior. Indeed, the problem of mixed colonies containing derivatives from multiple founder cells has already been identified[15] as bringing to question the safety of hiPSCs in future clinical trials. As such, a quantitative model for predicting the parameters required to optimise the occurrence of hiPSC colonies from a single founder cell would be particularly useful, and we intend to pursue this in a future work.

## ACKNOWLEDGMENTS

We are grateful to Newcastle University and European Community funding (IMI-STEMBANCC, IMI-EBISC and ERC # 614620) and to the School of Mathematics and Statistics of Newcastle University (Prof. R. Henderson) for providing partial financial support.

# SUPPLEMENTARY INFORMATION

# Dynamics of single human embryonic stem cells and their pairs: a quantitative analysis

L. E. Wadkin, L. F. Elliot, I. Neganova, N. G. Parker, V. Chichagova, G. Swan, A. Laude, M. Lako, and A. Shukurov

**Supplementary Table T1**

Table 1. Summary of parameters acquired for single cells cultured in the absence and presence of Cell Tracer. The entries represent the mean values with their standard errors and the spread within the sample given by the standard deviation (SD) of individual measurements around the mean. Migration velocities $v$ and step lengths $l_x$ and $l_y$ were calculated by averaging the displacements between images taken at 15 min intervals.

| Parameter and notation | | No Cell Tracer | | With Cell Tracer | |
|---|---|---|---|---|---|
| | | Average | SD | Average | SD |
| Number of cells | $N$ | 26 | – | 22 | – |
| Migration velocity ($\mu$m/hr) | $v$ | $22.7 \pm 0.6$ | 16.5 | $14.1 \pm 0.4$ | 12.2 |
| Step length in $x$ ($\mu$m) | $l_x$ | $3.5 \pm 0.1$ | 3.4 | $2.2 \pm 0.1$ | 2.4 |
| Step length in $y$ ($\mu$m) | $l_y$ | $3.5 \pm 0.1$ | 3.5 | $2.1 \pm 0.1$ | 2.5 |
| Time to first division (hr) | $t_{\text{div}}$ | $7 \pm 1$ | 5 | $10 \pm 2$ | 6.5 |

**Supplementary Table T2**

Table T2. Parameters characterising the migration of hESC pairs, both in the absence and in the presence of Cell Tracer. For each parameter, we present its mean value and standard deviation, as well as the spread given by the standard deviation of the individual measurements from the mean.

| Parameter and notation | | No Cell Tracer | | | | With Cell Tracer | |
|---|---|---|---|---|---|---|---|
| | | Type A | | Type B | | | |
| | | Average | SD | Average | SD | Average | SD |
| Number of pairs | $N$ | 10 | | 12 | | 18 | |
| Tracking time (hr) | | 23 | | 24.25 | | 31.5 | |
| Initial separation ($\mu$m) | $s_0$ | $17 \pm 1$ | 4 | $18 \pm 1$ | 5 | $18 \pm 1$ | 6 |
| Final separation ($\mu$m) | $s_f$ | $123 \pm 27$ | 87 | $42 \pm 9.$ | 32 | $19 \pm 2$ | 7 |
| Maximum separation ($\mu$m) | $s_{\max}$ | $141 \pm 30$ | 95 | $117 \pm 17$ | 58 | $49 \pm 2$ | 8 |
| Time to $s_{\max}$ (hr) | $T_{\max}$ | $16 \pm 2$ | 6 | $7 \pm 1$ | 4 | $7 \pm 2$ | 8 |
| Speed of each cell ($\mu$m/hr) | $v$ | $25.8 \pm 0.5$ | 20.4 | $24.8 \pm 0.4$ | 17.5 | $17.5 \pm 0.4$ | 14.0 |
| Pair centroid speed ($\mu$m/hr) | $v_{\text{pc}}$ | $18.7 \pm 0.5$ | 13.6 | $18.6 \pm 0.4$ | 12.4 | $14.1 \pm 0.4$ | 11.1 |
| Relative speed ($\mu$m/hr) | $v_r$ | $38 \pm 1$ | 28 | $33.3 \pm 0.8$ | 23.7 | $20.8 \pm 0.6$ | 17.1 |

**Supplementary Section S1: Directionality**

*Directionality,* (or the straightness index) is a simple and convenient parameter to quantify an isotropic random walk, as employed by Li *et al*[1]. The displacement of the cell at a time $t$, measured along the straight line from the starting point is $L_i = \sqrt{[x_i(t) - x_{i,0}]^2 + [y_i(t) - y_{i,0}]^2}$ (with $i$ the cell identifier) and the total distance traversed during the time $t$ is denoted $T_i$. The directionality of the cell migration is then defined as $\Delta_i = L_i/T_i$, and its values lie in the range $0 \leq \Delta_i \leq 1$. If the cell moves along a straight path, we have $T_i = L_i$, and the directionality has its maximum value, $\Delta_i = 1$. If, however, the cell follows a long and tortuous trajectory, then $T_i$ is much larger than $L_i$, and the directionality is low, $\Delta_i \approx 0$. Thus, the directionality quantifies how tangled and convoluted the cell's trajectory is. This quantity is closely related to the *tortuosity*, similarly characterising the shape of convoluted trajectories[2]. While the directionality may not the most useful characteristic of trajectories[3], we use it to retain comparability with earlier work on cell kinematics[3]. In particular, the directionality depends on the number of steps taken in the random walk. However, the unstained and stained cells move with similar correlation times, performing similar number of steps per unit time; this allows us to compare their trajectories in real time. It would not be difficult to describe the trajectories in terms of the number of random walk steps, but such a description would be less intuitive.

For a two-dimensional isotropic random walk, with steps of a length $l$, the average displacement from the starting point increases with the number of steps $N$ as $L_i = l\sqrt{N}$, where $N = t/\tau$ is the number of steps in time $t$. Meanwhile, the total distance traversed is $T_i = lN$. Then the average directionality of an isotropic random walk varies with time as[11]

$$\Delta_i \simeq \frac{1}{\sqrt{N}} = \sqrt{\frac{\tau}{t}}, \qquad (1)$$

decreasing towards zero as the number of steps $N$, or time $t$, increases. The reduction of the average directionality with time in inverse proportion to the square root of time elapsed since the start of the migration is a diagnostic property of an isotropic random walk. Note that Equation (1) gives the *averaged* directionality; the displacement and directionality for a single walker may deviate significantly due to the probabilistic nature of the walk.

To further confirm that the cell migration for unstained cells is consistent, on average, with the theory of isotropic random walk, we consider the average directionality (averaged over all 26 single hESCs) versus time, shown in Figure S1a. Up to around 7 hours there is a systematic decrease in the averaged directionality from unity to low values, in qualitative agreement with the random walk behaviour. To ascertain the functional form of this decay, the data is plotted on log-log axes (inset of Figure S1a). The prominent straight-line behaviour during this time indicates that the directionality decays as a power-law with time, and a straight-line least-squares fit (not constrained to go through any particular point) gives $\overline{\Delta}_i(t) = (0.50 \pm 0.02)t^{-0.44\pm0.04}$. The scaling with time is close to the $t^{-1/2}$ dependence characteristic of the isotropic random walk, Equation (1). Beyond 7 hours, the evolution of the average directionality changes its character and deviates from the $1/\sqrt{t}$ random walk behaviour; a similar deviation was noted in the plot of mean-square displacement versus time in Figure 3.

The averaged directionality in the presence of Cell Tracer, shown in Figure S1b, also indicates the systematic decrease over time, characteristic of an isotropic random walk. The least-squares fit is $\overline{\Delta}_i(t) = (0.61 \pm 0.05)t^{-0.50\pm0.04}$, which is also close to the $t^{-1/2}$ scaling characteristic of the diffusive motion.

The directionality of the unstained pair centroid motion, shown in Figure S2a, confirms that the pair as a whole can be described as an unbiased random walk at a good level of accuracy over the observation time range. The least-squares fit for unstained cells $\overline{\Delta}_i(t) = (0.49 \pm 0.02)t^{-0.42\pm0.02}$. The pair centroid motion of Cell Tracer stained cells is also consistent with a random walk: the least-squares fit is $\overline{\Delta}_i(t) = (0.54 \pm 0.06)t^{-0.56\pm0.06}$, shown in Figure S2b.

**Supplementary Figure S1**

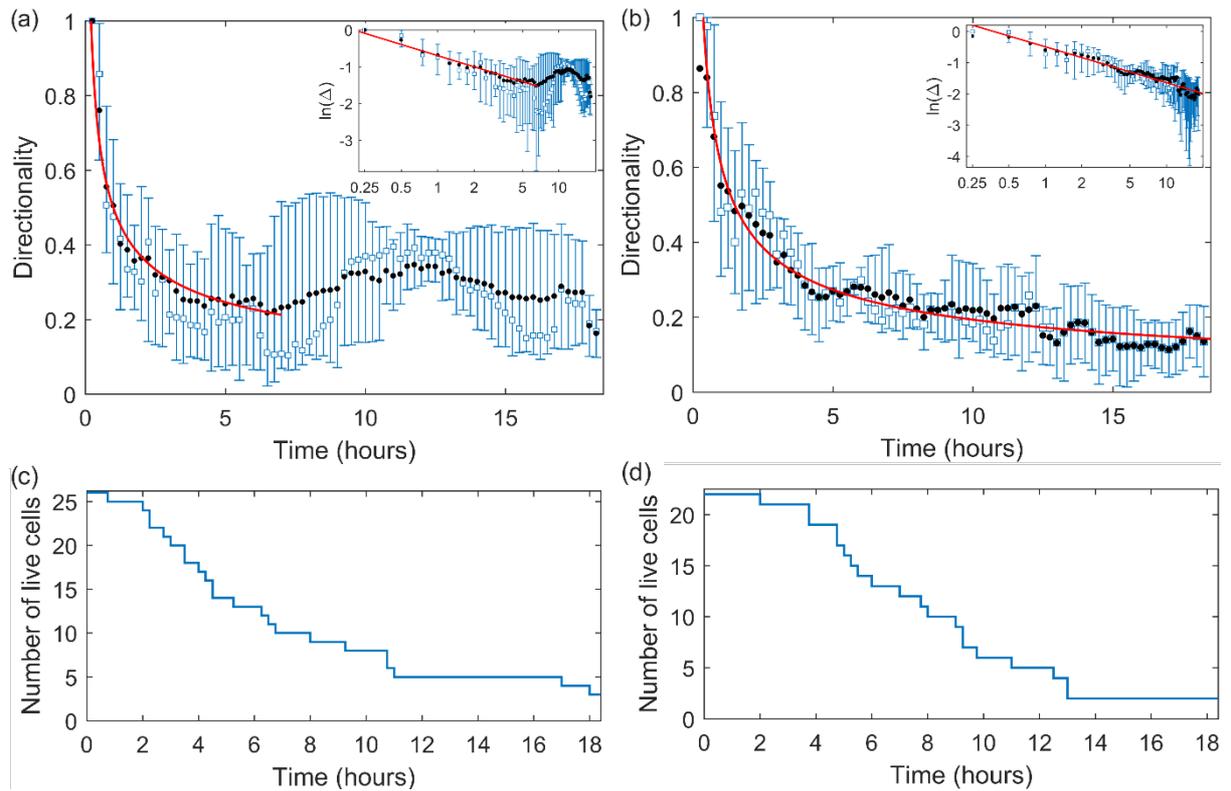

Supplementary Figure S1: Mean (black circles) and median (blue squares) directionality over time for the migration of single hESCs in (a) the absence of Cell Tracer, and (b) the presence of Cell Tracer. Insets show the data on natural logarithmic axes. Straight lines are a least-squares fit, applied to the whole time range in (b) and up to 7 hours in (a). These fits are (a) $\bar{\Delta}_\iota(t) = (0.50 \pm 0.02)t^{-0.44\pm0.04}$ and (b) $\bar{\Delta}_\iota(t) = (0.61 \pm 0.05)t^{-0.50\pm0.04}$. Error bars show the upper and lower quartiles. The number of live cells over time for unstained cells (c) and stained cells (d) to indicate the changing sample size. The sampling interval is every 15 minutes.

**Supplementary Figure S2**

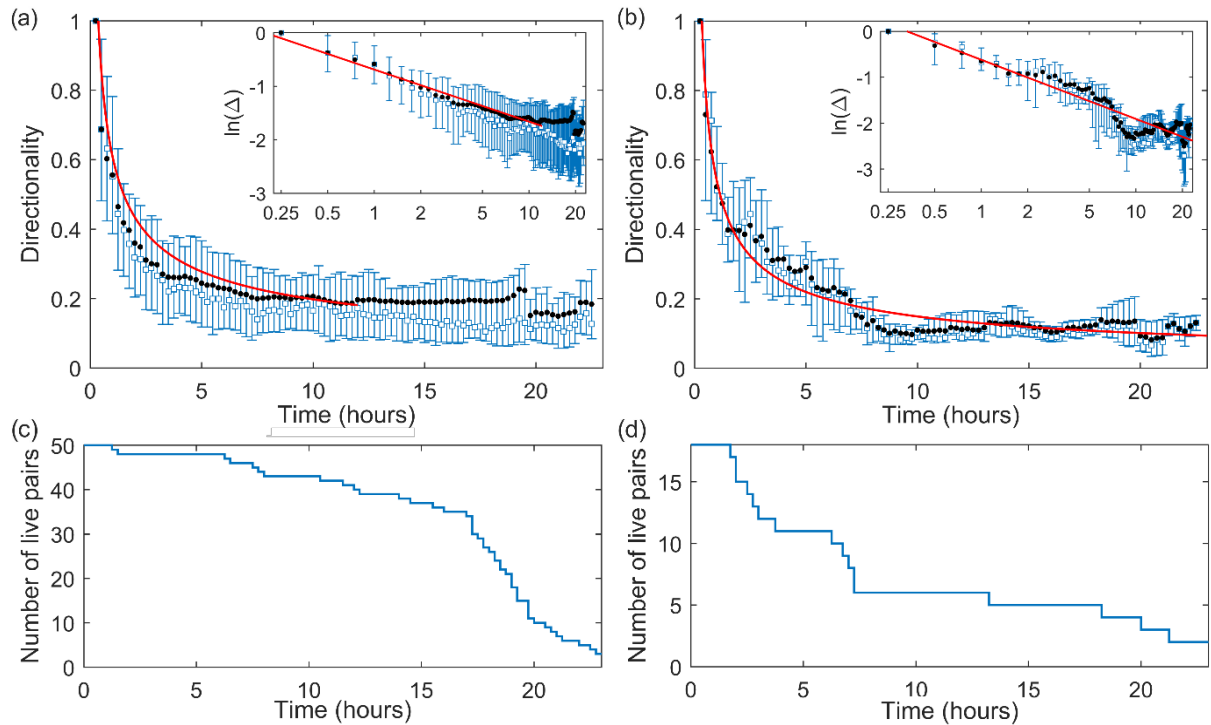

Supplementary Figure S2. Mean (black circles) and median (blue squares) directionality over time for the migration of pairs of hESCs in (a) the absence of Cell Tracer, and (b) the presence of Cell Tracer. Insets show the data on natural logarithmic axes. Straight lines are a least-squares fit, applied to the whole time range in (b) and up to 12 hours in (a). These fits are (a) $\bar{\Delta}_\iota(t) = (0.49 \pm 0.02)t^{-0.42 \pm 0.02}$ and (b) $\bar{\Delta}_\iota(t) = (0.54 \pm 0.06)t^{-0.56 \pm 0.06}$. Error bars show the upper and lower quartiles. The number of live cells over time for unstained cells (c) and stained cells (d) to indicate the changing sample size. The sampling interval is every 15 minutes.

**Supplementary Figure S3**

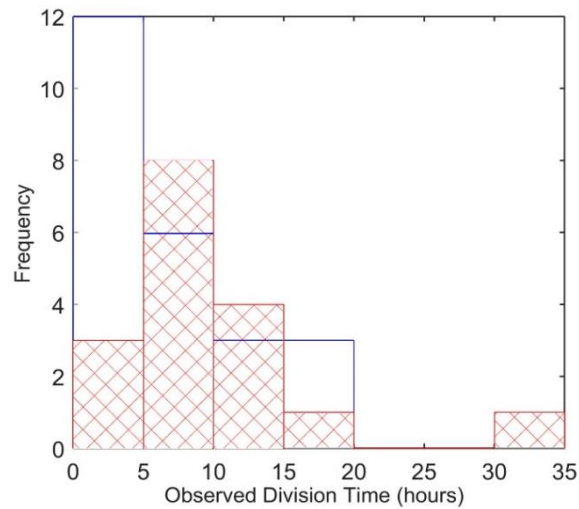

Supplementary Figure S3: Histograms of division times, with bin widths of 5 hours in each case, for the single hESCs in the absence (blue) and presence of Cell Tracer (cross hatched in red). The Kolmogorov–Smirnov two-sample test confirms that the two distributions are distinct, suggesting that the Cell Tracer treatment affects significantly the ability of the cells to divide. The Mann-Whitney U test also confirms the two distributions are distinct ($p < 0.05$).

**Supplementary Figure S4**

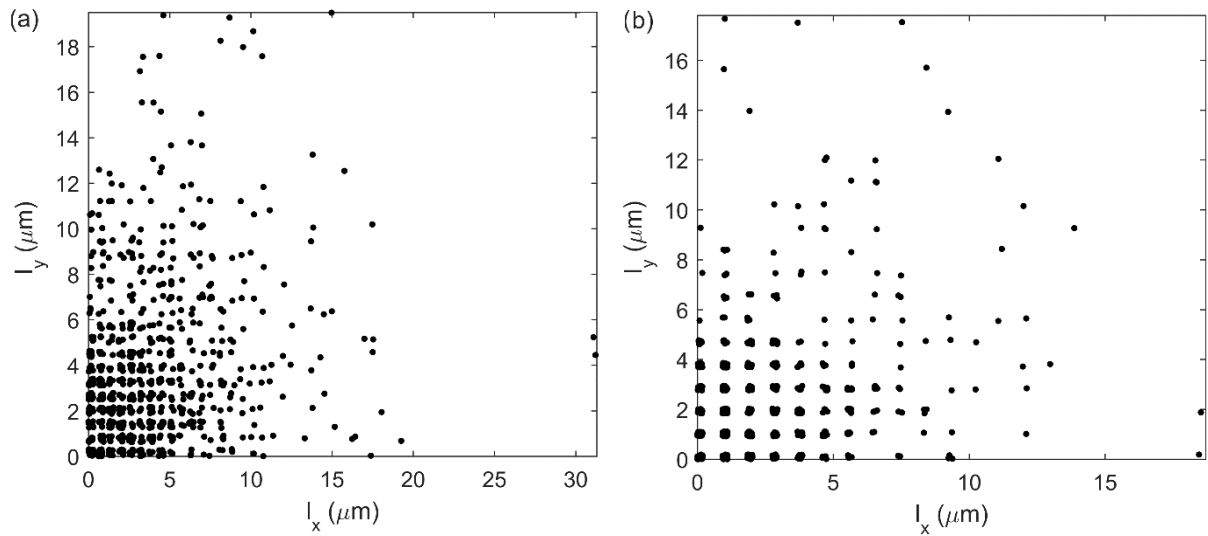

Supplementary Figure S4: The scatter plot of the step lengths $l_x$ and $l_y$ at each time frame (every 15 minutes) for cells (a) without Cell Tracer and (b) with Cell Tracer. Together with the low cross-correlation coefficient between the two variables discussed in the main text, the lack of any pronounced correlation between $l_x$ and $l_y$ [except perhaps the rare events with large values of $l_y$ in Panel (a)] suggests the isotropy of the random walk. According to the Kolmogorov-Smirnov and Mann-Whitney U tests, there is no evidence to distinguish between the distributions of $l_x$ and $l_y$. The Pearson product-moment correlation coefficient of $l_x$ and $l_y$ is as small as 0.22, confirming the steps in the $x$ and $y$ directions are uncorrelated.

**Supplementary Figure S5**

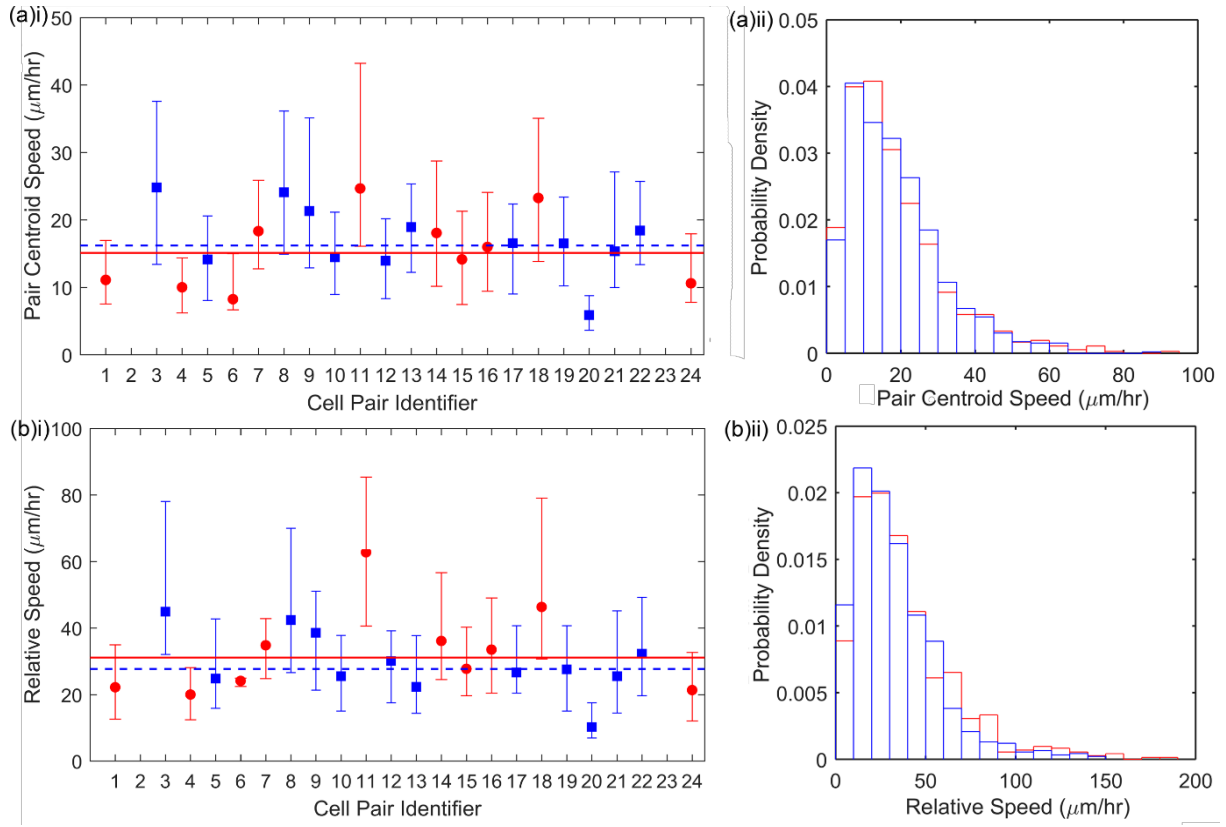

Supplementary Figure S5: (a) The speed of the pair centroid in the absence of Cell Tracer for the Type A (red) and Type B (blue) pairs: (i) the median speeds, with error bars representing the upper and lower quartiles, and (ii) the corresponding probability densities of the centroid speeds. Horizontal lines in (i) indicate the average across the entire category. (b): as in Panels (a) but for the relative speed within a pair. According to Kolmogorov–Smirnov and Mann-Whitney U tests the probability distributions for the Type A and Type B relative speeds are different.

**Supplementary Figure S6**

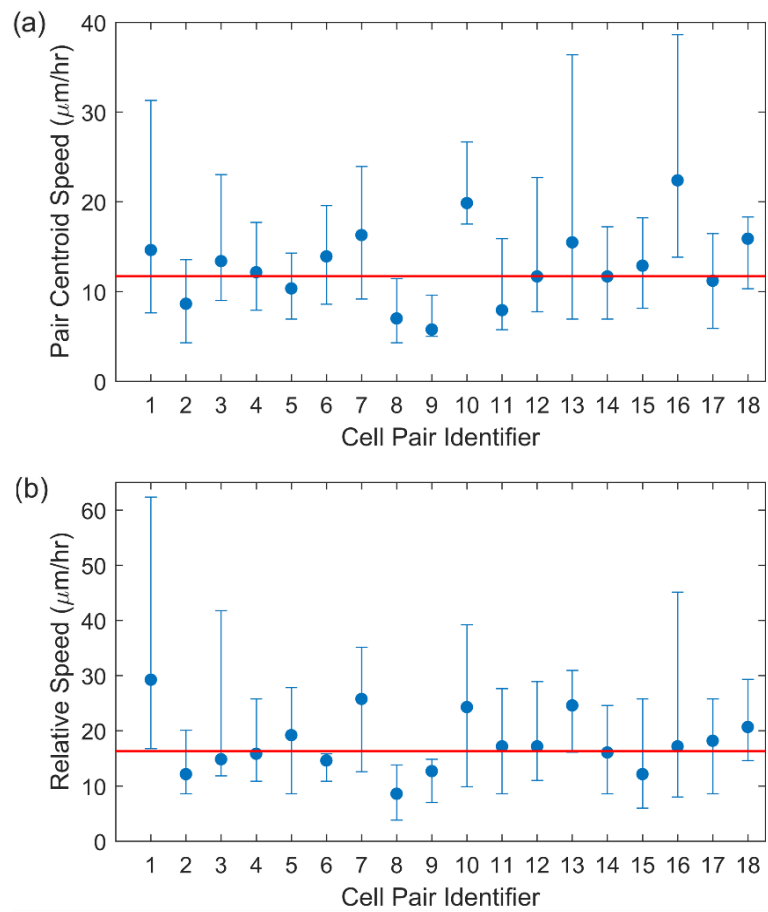

Supplementary Figure S6. (a) Median pair centroid speeds and (b) relative speeds of cell pairs in the presence of Cell Tracer.

**Supplementary Figure S7**

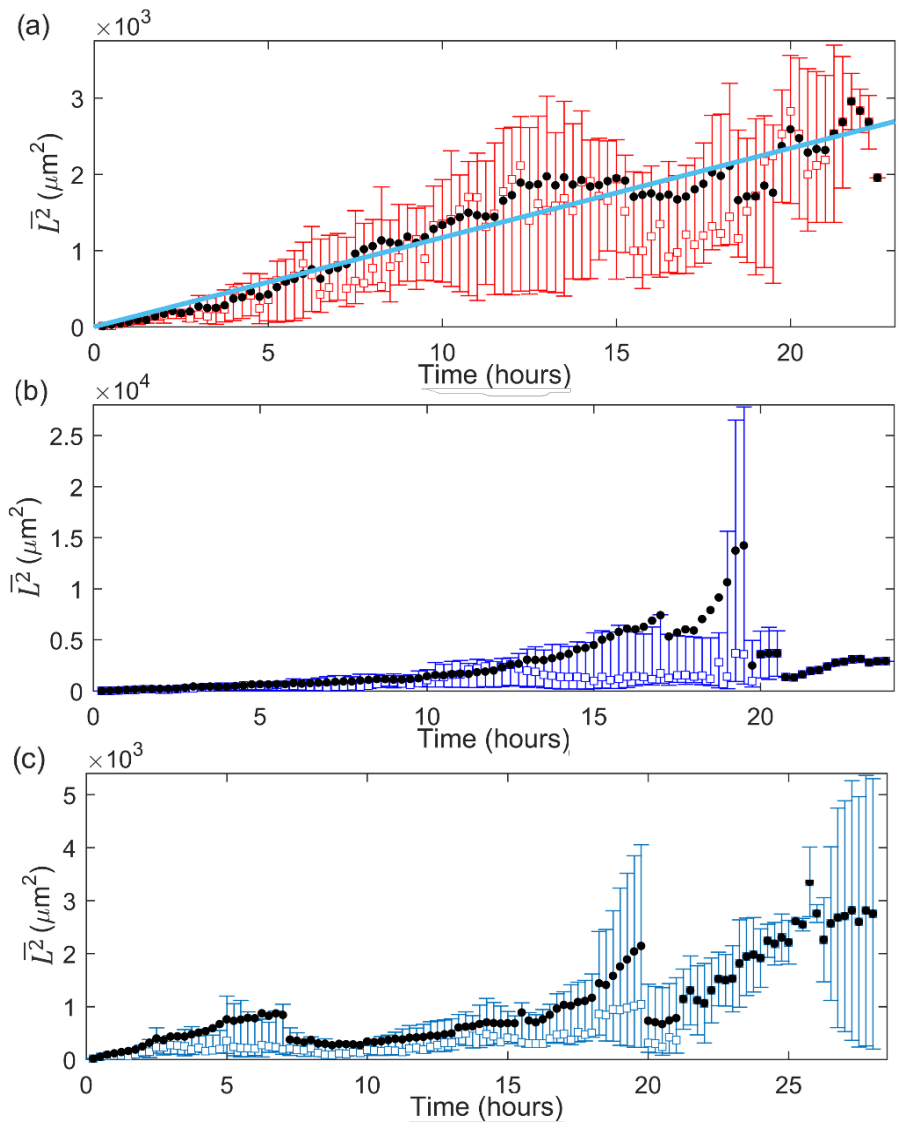

Supplementary Figure S7: The time evolution of the mean (black circles) and median (squares) centroid $\overline{L^2}$ for pairs of unstained cells of Type A (a) and unstained cells of Type B (b) and stained cells (c). The least-squares fit for Type A cells is $\overline{L^2} = 2Dt$ with $D = 58.45 \pm 1.8\ \mu m^2/hr$. Error bars show the upper and lower quartiles. The sampling interval is 15 minutes.

**Supplementary Section S3: Correlated Random Walk**

An alternative model for the migration of the cells is a correlated random walk, where the direction of a new step depends on the direction of the previous step, so that the migration retains a short-term memory of its direction[10]. We recall that the direction of movement is selected independently of the previous direction in the ordinary random walk. The migration remains diffusive over long time and length scales, but the diffusivity now depends on the mean value of the cosine of the angle $\theta$ between the two consecutive displacements, denoted here $\langle \cos \theta \rangle$:

$$D = \frac{\langle l^2 \rangle}{2\tau} + \frac{\langle l \rangle^2 \langle \cos \theta \rangle}{\tau(1 - \langle \cos \theta \rangle)},$$

where, as before, $\tau$ is the correlation time, $\langle l^2 \rangle$ and $\langle l \rangle$ are the mean squared length of the steps and their mean length, respectively, and angular brackets denote averaging[12]. For $\langle \cos \theta \rangle = 0$ (e.g., for $\theta$ uniformly distributed between 0 and $2\pi$), the standard expression (3) is recovered, with $\langle l^2 \rangle = \langle v^2 \rangle / \tau$. For $\langle \cos \theta \rangle \to 1$ (unidirectional motion), $D \to \infty$ signifying a non-diffusive motion.

For a correlated random walk, the relation of the correlation time to the diffusivity changes from $\tau = 2D/\langle v^2 \rangle = \langle l^2 \rangle/(2D)$, used in the main text, to

$$\tau = \frac{\langle l^2 \rangle/2 + \langle l \rangle^2 \langle \cos \theta \rangle/(1 - \langle \cos \theta \rangle)}{D}.$$

To assess possible importance of the short-time correlations, consider two cases that illustrate the range of possibilities, assuming $\langle l^2 \rangle = \langle l \rangle^2$ (e.g., a constant step length $l$). The diffusion coefficient obtained for an uncorrelated random walk is $D_0 = l^2/(2\tau)$. For $\langle \cos \theta \rangle = 1/2$, we obtain $D = 3D_0$, and the correlation time derived as $\tau = l^2/(2D)$ would is three times longer than its true value $\tau = l^2(1 + \langle \cos \theta \rangle)/[2D(1 - \langle \cos \theta \rangle)]$. Alternatively, for $\langle \cos \theta \rangle = -1/2$, we have $D = D_0/3$ and the correlation time inferred using the assumption of uncorrelated random walk is three times shorter than the true value.

We have not noticed any obvious signs that would suggest that the cell migration is better modelled as a correlated random walk. Since the correlation time of the cell migration is not affected much by the staining, the comparisons and conclusions discussed in the text are independent of this aspect of the random walk, if the staining only affects the parameters of the random walk (as we assume) rather than destroys or introduces any significant short-time correlations. However, this interesting question deserves further careful analysis.